\def\year{2020}\relax
\documentclass[letterpaper]{article} 
\usepackage{aaai20}  
\usepackage{times}  
\usepackage{helvet} 
\usepackage{courier}  
\usepackage[hyphens]{url}  
\usepackage{graphicx} 
\usepackage{graphicx}  
\usepackage{subcaption}
\usepackage{xcolor}
\usepackage{multirow}
\usepackage{tabularx}

\newcommand{\EC}{\textit{ExpCorrectFirst}}
\newcommand{\EW}{\textit{ExpWrongFirst}}
\newcommand{\NC}{\textit{NovCorrectFirst}}
\newcommand{\NW}{\textit{NovWrongFirst}}
\newcommand{\nov}{\textit{novice}}
\newcommand{\expert}{\textit{experienced}}
\newcommand{\corr}{\textit{correct first}}
\newcommand{\wro}{\textit{wrong first}}

\newcommand{\task}{\textit{arthropod review task}}

\title{The Role of Domain Expertise in User Trust and the Impact of First Impressions with Intelligent Systems}

\author{
\Large \textbf{Mahsan Nourani\textsuperscript{\rm1}, Joanie T. King\textsuperscript{\rm2}, Eric D. Ragan\textsuperscript{\rm1}}\\
\textsuperscript{\rm1} University of Florida, Gainesville, Florida \\
\textsuperscript{\rm2} Texas A\&M University, College Station, Texas \\
\textsuperscript{} mahsannourani@ufl.edu, joanie\_king@tamu.edu, eragan@ufl.edu}

\begin{document}

\maketitle

\begin{abstract}
Domain-specific intelligent systems are meant to help system users in their decision-making process.
Many systems aim to simultaneously support different users with varying levels of domain expertise, but prior domain knowledge can affect user trust and confidence in detecting system errors.
While it is also known that user trust can be influenced by first impressions with intelligent systems, our research explores the relationship between ordering bias and domain expertise when encountering errors in intelligent systems.
In this paper, we present a controlled user study to explore the role of domain knowledge in establishing trust and susceptibility to the influence of first impressions on user trust.
Participants reviewed an explainable image classifier with a constant accuracy and two different orders of observing system errors (observing errors in the beginning of usage vs. in the end).
Our findings indicate that encountering errors early-on can cause negative first impressions for domain experts, negatively impacting their trust over the course of interactions.
However, encountering correct outputs early helps more knowledgable users to dynamically adjust their trust based on their observations of system performance.
In contrast, novice users suffer from over-reliance due to their lack of proper knowledge to detect errors.

\end{abstract}

\section{Introduction}
\label{sec:Introduction}

System designers and practitioners incorporate machine learning and artificial intelligence (ML/AI) models to help end-users achieve their goals and make decisions.
Intelligent systems are used across a wide variety of domains, such as medical diagnosis assistance~\cite{goyal2018medical,bussone2015role}, cybersecurity monitoring~\cite{goyal2019semantic}, and criminal justice~\cite{rudin2018optimized,berk2015machine}.
The intended end users of such systems often possess different levels of background domain knowledge.
For instance, medical decision support systems incorporate AI/ML approaches to help with automated diagnoses for diseases.
While doctors and medical practitioners can use these systems to make a diagnosis or verify it, patients may use similar systems to input their symptoms for an early diagnosis. 

Previous research has demonstrated that domain expertise and user-reliance on intelligent systems are related.
For example, Bussone et al.~\shortcite{bussone2015role} have shown that little or no domain knowledge can cause over-reliance on the system and automation.
It has also been found that preexisting knowledge of an automated system can influence user's initial trust on the system~\cite{hoff2015trust}.
We can conclude that domain experience plays an important role on when users decide to trust a system and when not.
To encourage a more trustworthy intelligent system, researchers suggest incorporating explanatory techniques to improve transparency and help users understand how the model is making its predictions.~\cite{hoff2015trust,doshi2017towards}.

Though improving model transparency is viewed as a partial solution to the trust problem, users normally develop a sense of a system's accuracy through their own experiences and observations using the system over time.
Thus, when beginning to use the system, and before fully developing an appropriate mental model, early impressions of the system can affect how users perceive the system's accuracy~\cite{nourani2020investigating}.
Since there is almost no AI/ML algorithm with 100\% accuracy in meaningful real-world contexts, we expect all intelligent systems will eventually make errors---but \textit{when} users observe the errors is crucial.

As users interact more with an automated system over time, their trust evolves based on user observations and experiences~\cite{hoffman2013trust}.
We also might expect a different initial trust from users with domain experience while their ability in detecting errors can also affect how their trust changes---as also suggested by~\cite{merritt2015well}.

Motivated by these challenges, we designed an experiment to explore how users with different levels of domain expertise develop trust over time as the observed accuracy changes.
Incorporating a simulated image classification task in the entomology domain, we recruited domain-experienced and novice participants in an online study.
To study the relationship between domain knowledge and first impressions, we defined two extreme scenarios to control when users experience system errors: 1) in the beginning and 2) at the end of the usage.
Each user reviewed \textit{the same} set of arthropod images---with their associated labels and explanations---while the order of the set was determined by one of the two assigned scenarios in a between-subjects setting.
We measured and compared trust and its calibration over time for the novice and experienced participants based on their initial impressions of the system, as well as their perception of the overall system accuracy.
Our results provide novel and significant findings of the importance of domain knowledge in the formation of first impressions and experience with the system over time.

\section{Related Work}
\label{sec:RelatedWork}

Social and psychological researchers have been studying human trust for many years.
Although there is not one agreed upon definition of trust in this area, human-human trust is commonly based on believing that the trustee will do what is expected~\cite{good2000individuals}.
Similarly, trust in automation is a user's ability to rely on and predict the results from the automated system.
Similar to human-human trust, once human-machine trust is lost, it is hard to reestablish it~\cite{hoffman2013trust}. 
However, research has shown that humans are more forgiving towards humans than machines when their invested trust is violated~\cite{de2012world}, which highlights the importance of maintaining trust in automation.

Researchers have looked into different modes of trust in automated and intelligent systems.
For instance, Merrit et al.~\shortcite{merritt2015well} examined trust calibrations (users' adaptation of trust over time) and its outcomes on task-performance and error detection.
One of the major design decisions that is mutually accepted in the research community is model transparency through explanatory systems and intelligent user interfaces~\cite{hoffman2018explaining}.
Papenmier et al.~\shortcite{papenmeier2019model} studied the interplay between model accuracy and explanation fidelity, and how they affect user trust in intelligent systems.
Their results show that model accuracy plays a more important role on user trust than explainability.
They also found that users cannot be tricked into trusting a bad classifier when the system provides high fidelity explanations.
Our work is similar to their work in that we explore user trust through a low accuracy classifier system with high fidelity explanations.
However, we are looking into whether a user's domain knowledge can affect their perception of the system accuracy.
In other relevant work, Yu et al.~\shortcite{yu2017trust} studied changes of trust over time based on different levels of model accuracy through a decision-support system, targeting novice users.
They found that with lower overall accuracy, trust tends to decrease over time.
Our research also studies the changes of trust over time, while we focus on how first impressions can affect this change, specifically with domain experience.

Different factors (prior and during interactions) can affect user trust and reliance on intelligent systems~\cite{hoff2015trust}.
These factors can make it more challenging to design intelligent systems and agents.
One known challenge is when users tend to over-trust and over-rely on the system predictions, heavily depending their decisions on the system outputs (also known as \textit{automation bias})~\cite{mosier1999automation,alberdi2005automation}.
Previous research demonstrate that novice users can suffer from this problem~\cite{mosier1999automation,bussone2015role}.
On the contrary, mistrust and distrust can cause users to underestimate the system and rely on themselves, eventually causing them to stop using the automated system in the future.
For example, getting people to provide feedback to an intelligent system to fix errors can amplify user mistrust in the system~\cite{honeycutt2020solicit}.
More related to this paper, Nourani et al.~\shortcite{nourani2020don} found that after observing system's weakly-justified predictions, users tend to disagree with the system even when it is right.
In this paper, we explored the interplay of domain expertise and observed performance on user reliance behaviours.

A number of researchers in human-computer interaction have researched how domain expertise can affect user behaviours with intelligent systems.
For example, Zhang et al.~\shortcite{zhang2020effect} studied how users' trust calibration can be affected by knowing that their domain knowledge is higher than the model's.
Although they raise an important question, their results were inconclusive.
The resulting studies have been focused on different domains, such as medical~\cite{bussone2015role,vaidyanathan2014recurrence,cai2019hello}, data science~\cite{kaur2020interpreting}, visual analytics~\cite{dasgupta2016familiarity}, and aviation~\cite{mosier1999automation}.
In this paper, unlike the previous work, we study how domain expertise can affect impression formation and trust calibration.
It is important to bear in mind that novice and expert terminology is task and domain dependent and might vary from one system to another.
For example, some researchers define novice users as students or those who have a ground-level of knowledge in the domain~\cite{bussone2015role,mosier1999automation}, while many times terms as novice or lay users are referring to the general public with close to zero knowledge in the domain\cite{doshi2017towards}.

Trust is subjective by nature and therefore challenging to quantify.
Several methods have been suggested to measure trust in intelligent systems~\cite{mohseni2018survey}.
There are many suggested \textit{trust in automation} questionnaires such as~\cite{hoffman2018metrics} that can be used to measure trust explicitly.
Some of the implicit trust measurements include checking user agreement with wrong system outputs~\cite{papenmeier2019model,nourani2020don}; repeatedly asking for trust ratings~\cite{yu2017trust}; and measuring user perception of system accuracy as an indication of user trust~\cite{yin2019understanding,nourani2019effects}.
In this paper, we measure trust through both implicit and explicit measures.

In this paper, we study first impression formation based on domain expertise.
Previous research on first impressions has shown that human's early observations and judgements can bias and affect their behaviours towards people~\cite{zebrowitz2017first,fourakis2020matters}, systems~\cite{nourani2020investigating}, and/or agents~\cite{petrak2019let,desai2013impact}.
However, to the best of our knowledge, there has been little focus on how the ordering of user experiences with different model outputs can affect user trust.
Without focusing on the order, Dietvorst et al.~\shortcite{dietvorst2015algorithm} found that users tend to avoid an algorithm and favor humans over systems once they observed the system made an error.
In our own previous work, we found evidence that positive and negative first impressions can affect user reliance and mental models of an intelligent system.
However, the prior study did not account for relationships including domain expertise and changes in trust over time~\shortcite{nourani2020investigating}, which serves as the focus for our new study of first impressions in the current paper.

\section{Experiment}
\label{sec:Experiment}
We conducted a user study with a simulated multi-class image classification scenario to understand how domain knowledge and order of observing system errors can affect user trust.
In this section, we discuss the study design, goals, and participants in more detail.

\subsection{Research Goals and Hypotheses}

The primary motivation of this study was to understand how first impressions of an intelligent system can affect user trust, and whether and how domain expertise can help bypass the influences of these first impressions.
We focused on systems with local outputs and explanations, i.e., systems that show one output at a time to their users -- e.g., ~\cite{ribeiro2016should} --  rather than intelligent systems at a global scope where users see a representation of how the model works on a high level -- e.g., \cite{hohman2019s}.
Considering these systems, our goal was to understand how user domain expertise affects the formation of first impressions, changes of trust over time, and estimation of system accuracy.
To address this research question, we summarized the following set of hypotheses:

\begin{itemize}
    \setlength\itemsep{0.05cm}
    \item \textbf{H1}: Ordering bias only affects first impression formation in users with domain expertise, whereas novice users are more prone to automation bias due to having constantly high trust on the system.
    \item \textbf{H2}: Users with domain expertise and positive first impressions have a higher overall trust on the system compared to those with negative first impressions.
    \item \textbf{H3}: Users with domain expertise and positive first impressions will adjust their trust over time based on their observation of system errors, while those with negative first impressions will continue mistrusting the system, regardless of their observation of the system performance.
\end{itemize}

\begin{figure}[!t]
\centering
  \captionsetup{belowskip=-15pt}
  \includegraphics[width=0.99\columnwidth]{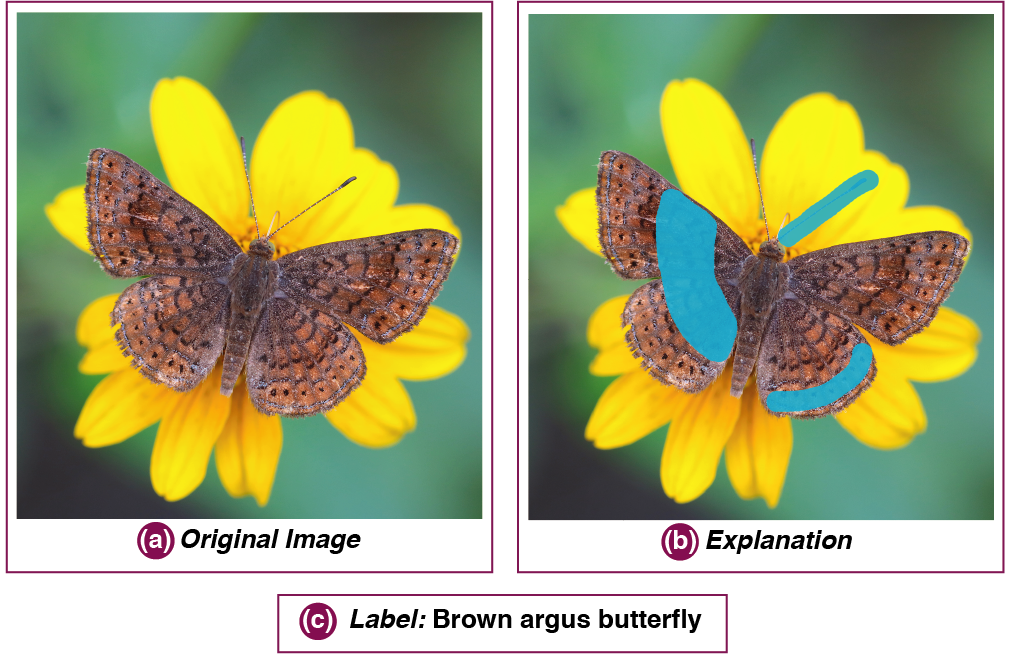}
  \caption{An example of what participants observed in the study. (a) the original image (which is a \textit{Calephelis} sp.); (b) a blue saliency map to explain which regions of the image were used by an expert to determine the name and species of the arthropod; (c) the image classification label, which in this case is \textit{incorrect}.} ~\label{fig:example}
\end{figure}

To test these hypotheses, we controlled two different orders of presenting system outputs:
(1) Participants observed all the correct predictions in the beginning and all the mis-predictions at the end (i.e., a positive first impression) and (2) observations follow the opposite order (i.e., a negative first impression).
Note that the only difference in these conditions was the order of presenting the output, while the accuracy and observed trials remained the same.
Figure~\ref{fig:example} shows an example of an image with its corresponding label and explanation.
\begin{figure}[!t]
    \begin{subfigure}{0.49\columnwidth}
        \centering
        {\includegraphics[width=\columnwidth]{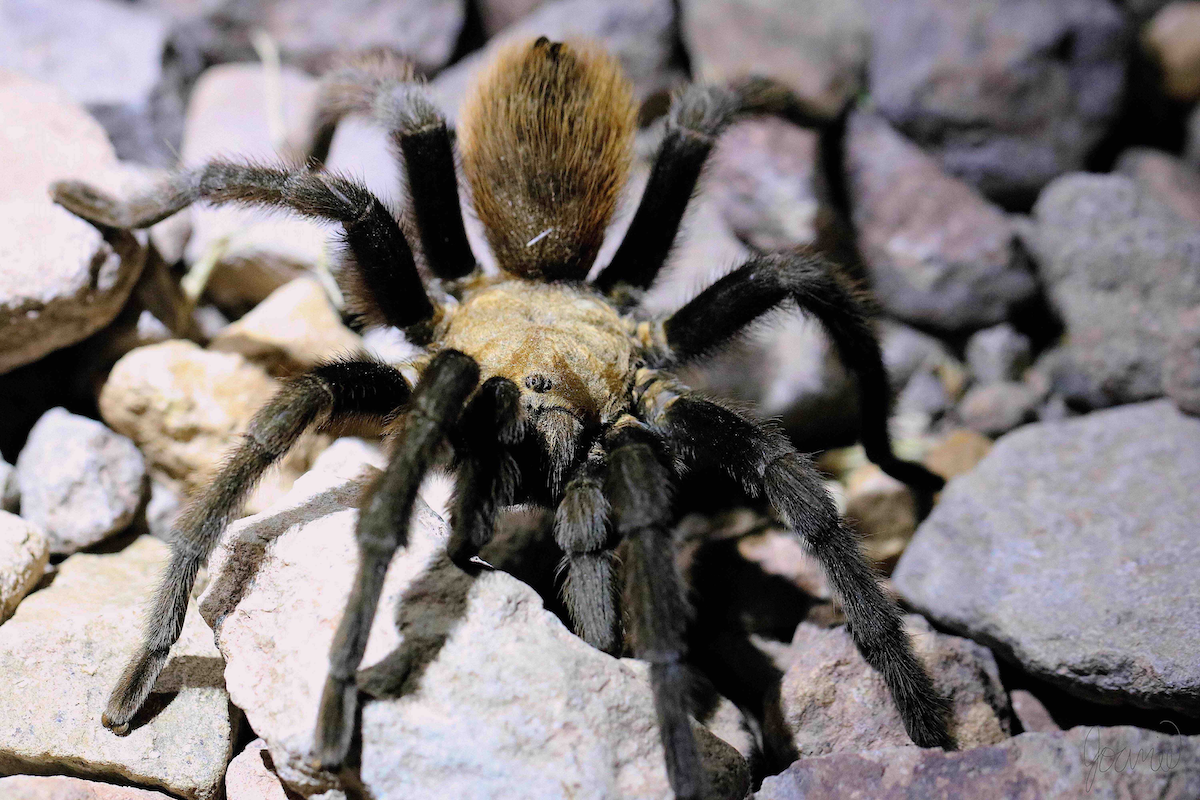}}
        \captionsetup{justification=centering}
        \caption{Desert Tarantula}
        \vspace{0.35cm}
        \label{fig:easy}
    \end{subfigure}
    \begin{subfigure}{0.49\columnwidth}
        \centering
        {\includegraphics[width=\columnwidth]{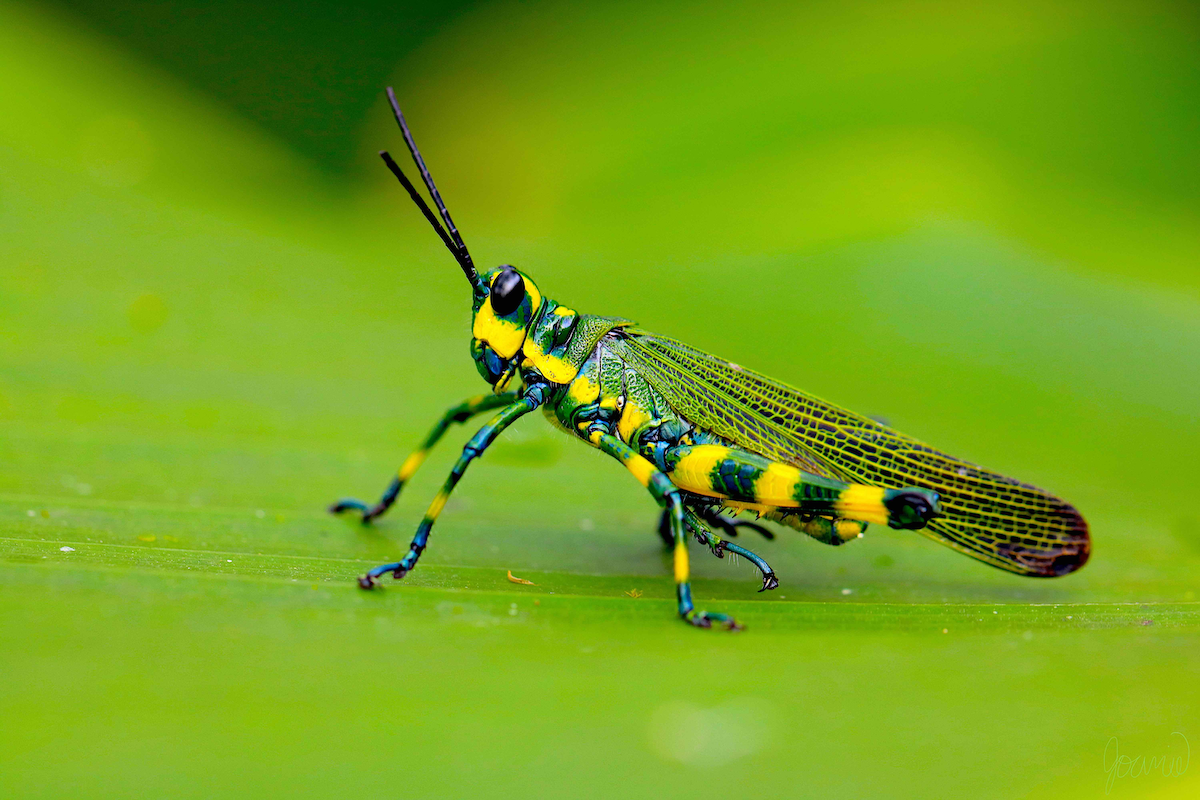}}
        \captionsetup{justification=centering}
        \caption{Ecuadorian Lubber Grasshopper}
        \label{fig:easy2}
    \end{subfigure}
    \begin{subfigure}{0.49\columnwidth}
        \centering
        {\includegraphics[width=\columnwidth]{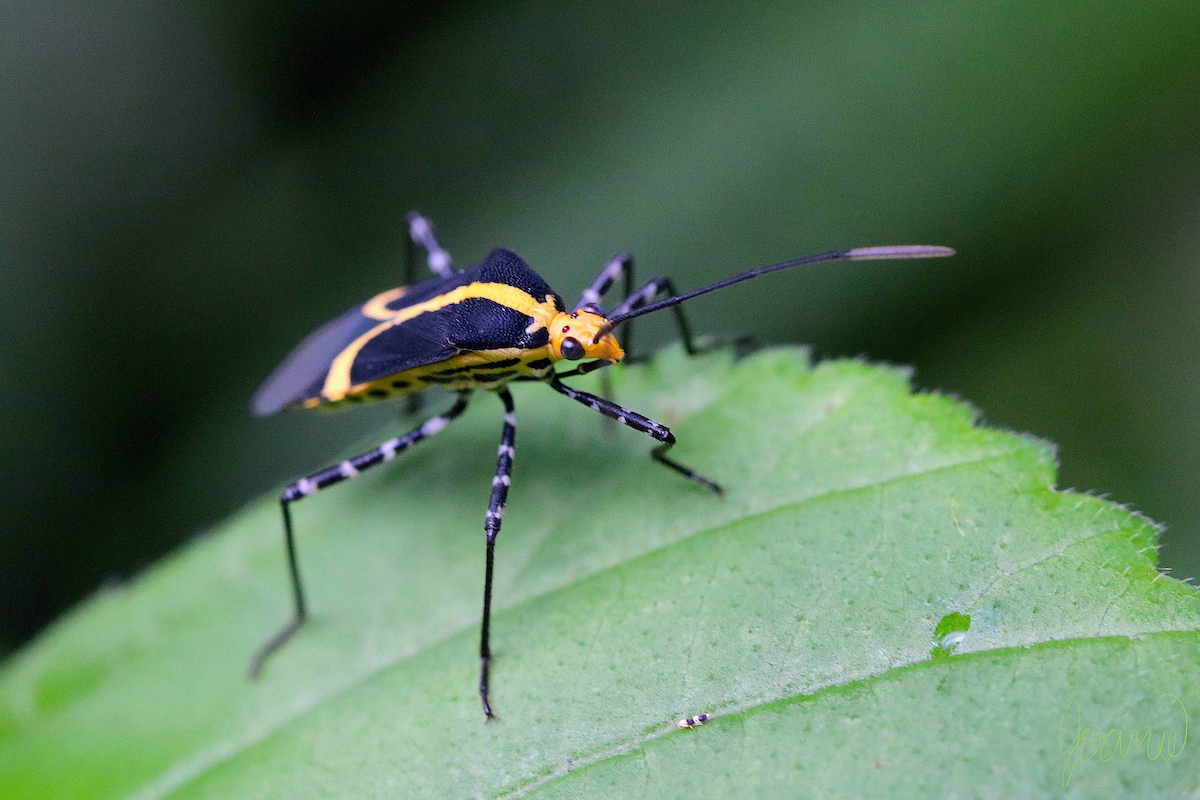}}
        \captionsetup{justification=centering}
        \caption{Rhopalid \\(scentless plant bug)}
        \label{fig:hard}
    \end{subfigure}
    \begin{subfigure}{0.49\columnwidth}
        \centering
        {\includegraphics[width=\columnwidth]{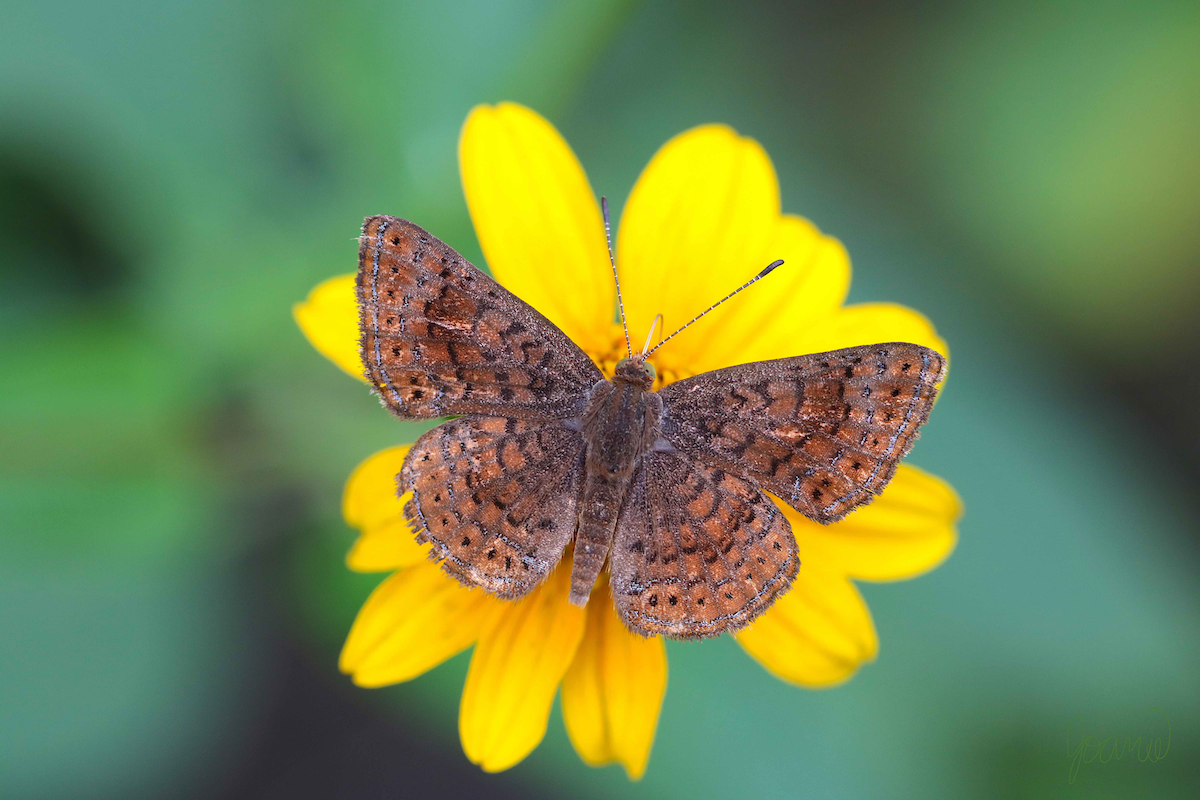}}
        \captionsetup{justification=centering}
        \caption{Metalmark Butterfly \\ (\textit{Calephelis} sp.)}
        \label{fig:hard2}
    \end{subfigure}
    \caption{Four examples of raw images from the study dataset. (a) and (b) are examples of easy-to-detect images while (c) and (d) are examples of hard-to-detect images.}
    \label{fig:rawImages}
\end{figure}

\subsection{Experimental Design}
To test our hypotheses, we designed a user study where participants were asked to review a set of images from a multi-class simulated classification scenario.
Based on our research questions and goals, we sought an image classification domain where background knowledge is not a requirement, while having it could help a lot in completing the task.
We chose entomology---the study of insects and other terrestrial arthropods---as a domain where novice users can partially identify system errors, but domain experts are expected to excel at this task.
The \task{} consisted of a set of 40 trials with a classification accuracy of 47.5\%, and was designed to allow the participants to experience and use the classification system over time in order to measure their level of trust.

We defined two independent variables for the study: \textit{domain knowledge} and \textit{order of observing correct outputs}.
First, \textit{domain knowledge} refers to a user's level of familiarity with and knowledge of entomology, for which we defined two levels: \nov{} and \expert{}.
Second, we controlled the \textit{order of observing outputs} in two different manners.
For the \corr{} level, all the correctly-classified outputs were observed in the beginning while all the mis-classifications were shown afterwards.
The reverse order was provided for the \wro{} level.

The study followed a 2x2 between-subjects design, where each participant from \nov{} and \expert{} group completed and observed the trials in one of the two defined orders.
Since the subjects were exposed to the same set of trials, only with a different order, we incorporated a between-subjects design over a within-subjects design in order to avoid biases and learning effects.

\subsection{Dataset}
\label{sec:dataset}
For the purposes of the study, the experiment's data used 40 high-quality macro images of different arthropods, photographed by an entomologist on the research team.
Different regions in the world have bug species that are specific to each area and might not be found in other places.
Since we were running the study in the US and our target entomology participants were mostly familiar with the arthropods in this country, all the selected arthropod images were from arthropod families that are found in the United States.
Figure~\ref{fig:rawImages} shows examples of the raw images used in the study.

As our study was designed for both \nov{} and \expert{} users, we designed the image set to contain a mix of both easy-to-detect and hard-to-detect arthropods in order to make the task fair for the \nov{}\textit{s} as well. 
After selecting the images for the study, our expert entomologist generated a textual classification label for each image.
The label contained the name of the arthropod and, in some cases, the family and species of the arthropod in brackets.
For each image, our expert also created high-fidelity explanations in the form of saliency maps on top of the image.
These saliency maps were chosen as portions of the image that the expert would use herself to detect the bug in the image.
However, to address our goals and hypotheses of the study, we selected the classification accuracy of the simulated system as 47.5\%, i.e., 19 images included correct labels and 21 images included false labels.

\subsection{Participants}

We recruited a total of 116 participants for this study, with 48 females, 61 males, and 7 others (non-binary, non-listed, or unknown).
For the purposes of this study, we distinguish two groups of participants: 1) people who had \textit{at least} 1 year of university coursework in entomology (i.e., the \expert{}), and 2) people with little or no familiarity with entomology (i.e., the \nov{}s).
These two groups were recruited separately.
The \nov{} participants were recruited from undergraduate and graduate level university students, most of whom studying in computing majors.
The \expert{} subjects were university students and practitioners in entomology or related fields.
Among these participants, 71.23\% held or were pursing a graduate degree. 

To help verify participants were considered in the appropriate group for expertise level, participants self-reported their level of familiarity with entomology as well as their occupation or major.
Familiarity was measured through a seven-point Likert scale from 1 to 7 for \textit{no knowledge} to \textit{expert}, respectively.
Since this self-reported measure is subjective, novices might overestimate their knowledge, whereas experts might underestimate it~\cite{aqueveque2018ignorant,dunning2011dunning}.
A two-way factorial ANOVA found significant differences between these groups, with $F(1,107) = 712.99$, $p<0.001$, showing the domain-experienced group significantly rated their familiarity higher than novices.

\begin{figure}[!t]
\centering
  \captionsetup{belowskip=-10pt}
  \includegraphics[width=0.99\columnwidth]{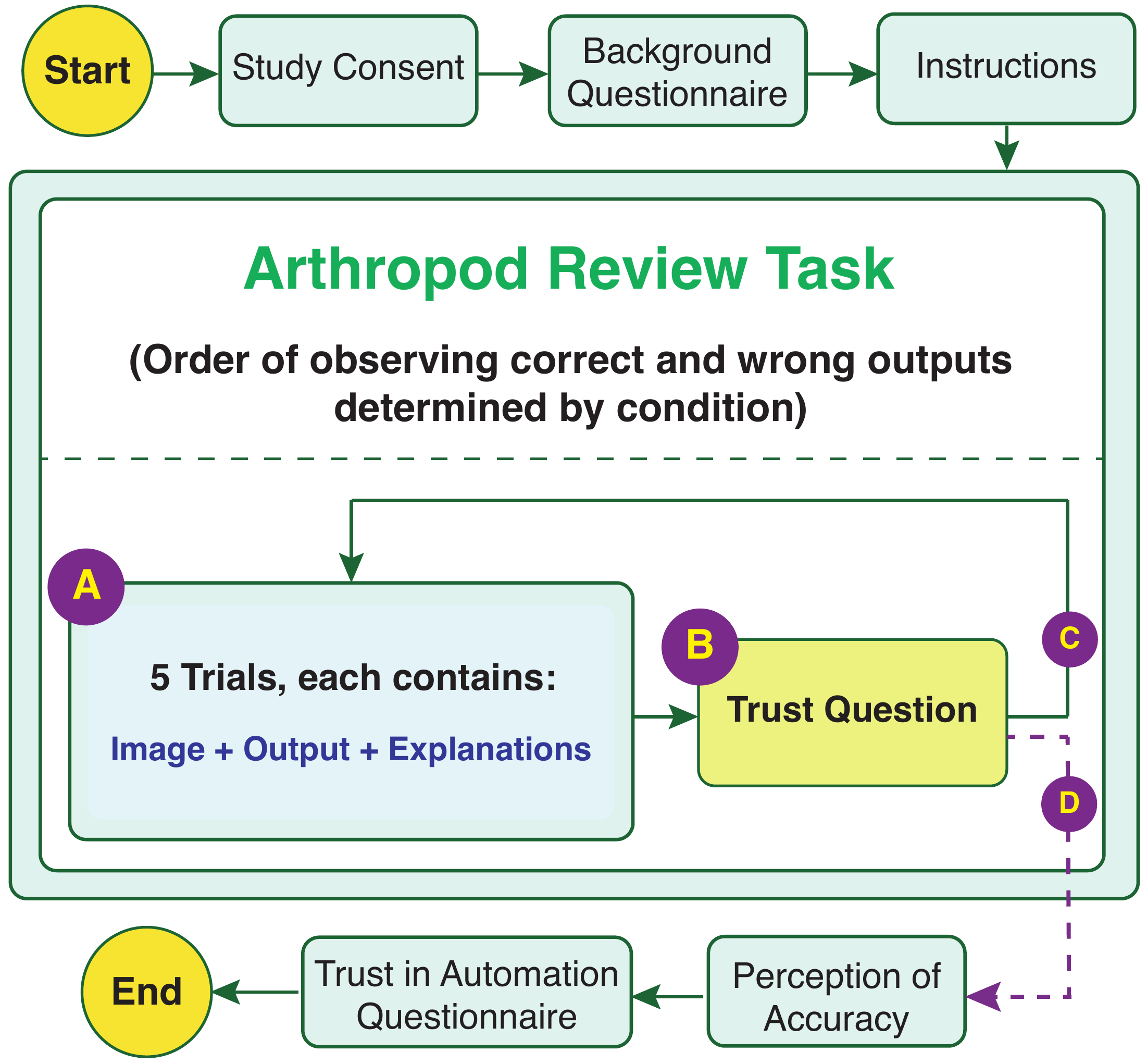}
  \caption{An overview of the study procedure. The Arthropod Review Task starts with \textit{(A)}, consisting of 5 trials, and continues to a trust question \textit{(B)}. By following \textit{path (C)}, participants iterate through \textit{(A)} and \textit{(B)} seven more times. After answering to the trust question for the $8^{th}$ time, subjects continue to the post-study questionnaire through \textit{path (D)}.} ~\label{fig:procedure}
\end{figure}

\subsection{Study Procedure and Measures}

The user study was conducted online through a custom web application and took roughly 20 minutes.
Participants were asked to complete the study in a single session using a preferred web browser on a desktop computer. 
The study was approved by the organization's Institutional Review Board (IRB).
Figure~\ref{fig:procedure} shows the overall procedure of the study.

The participants filled out a background questionnaire about demographic information, education, occupation, and familiarity with machine learning and entomology.
They were shown instructions about the study and the task.
After reviewing the instructions, participants started the \task{}, where they reviewed 40 trials.
Each trial consisted of: 1) an image of an arthropod, 2) a textual label for the classification of the image (which might also include the family and species of the bug), and 3) a feature explanation for the classification in form of a saliency map (as described in section~\ref{sec:dataset}).

For each trial, participants were required to rate their agreement with two statements on a 5-point Likert scale, as seen below.
In order to answer these questions, participants were advised to use their best judgement for identifying the arthropod in each trial, and in case the bug was unfamiliar, they were advised to refer to the provided explanations.

\begin{enumerate}
    \setlength\itemsep{0.005cm}
    \item \textit{I believe the highlighted explanation is appropriate with regards to the system answer.}
    \item \textit{I am confident that the system answer is correct.}
\end{enumerate}

These questions were meant to focus user attention to the label and explanation of each image before moving on and to build an understanding of how the classifier works.
After every five trials, participants were asked to report their level of trust on the system based on their observations up to that point.
Trust was rated on a 7-point Likert scale from 1 (\textit{distrust}) to 7 (\textit{trust}).

After the \task{}, participants answered a questionnaire on trust in explainable AI~\cite{hoffman2018metrics} and estimated the system accuracy in percent.
They also answered two free-response questions, asking them to explain how the system accuracy and their trust changed over time.

\section{Results}

\begin{table}[!t]
    \resizebox{\columnwidth}{!}{
    \begin{tabular}{|c|c|}
         \hline
         \rule{0pt}{2ex}
         \multirow{8}{*}{\textbf{(a) Average Trust Rating}} & Main Effect\\ \cline{2-2} \rule{0pt}{2.5ex} 
         & \multicolumn{1}{r|}{\textit{Domain Knowledge}: $F(1,107) = 39.07, p < 0.001$ \textbf{\textcolor{red}{*}}} \\ \rule{0pt}{2.5ex} 
         & \multicolumn{1}{r|}{\textit{Order of Trials}:  $F(1,107) = 17.66, p < 0.001$ \textbf{\textcolor{red}{*}}} \\\rule{0pt}{2.5ex} 
         & \multicolumn{1}{r|}{\textit{Interaction Effect}: $F(1,107) = 26.14, p < 0.001$ \textbf{\textcolor{red}{*}}} \\  \cline{2-2} \rule{0pt}{2.5ex} 
         & Post-Hoc Test \\ \cline{2-2} \rule{0pt}{2.5ex} 
         & \multicolumn{1}{r|}{\textbf{\EC{}} vs. \EW{} ($p<0.001$) \textbf{\textcolor{red}{*}}} \\ \rule{0pt}{2.5ex} 
         & \multicolumn{1}{r|}{\textbf{\NW{}} vs. \EW{} ($p<0.001$) \textbf{\textcolor{red}{*}}} \\ \rule{0pt}{2.5ex} 
         & \multicolumn{1}{r|}{\textbf{\NC{}} vs. \EW{} ($p<0.001$) \textbf{\textcolor{red}{*}}} \\ 

         \hline
         \hline
         \rule{0pt}{2ex}
         \multirow{8}{*}{\textbf{(b) Change of Trust Rating}} & Main Effect\\ \cline{2-2} \rule{0pt}{2.5ex} 
         & \multicolumn{1}{r|}{\textit{Domain Knowledge}: $F(1,107) = 17.58, p < 0.001$ \textbf{\textcolor{red}{*}}} \\ \rule{0pt}{2.5ex} 
         & \multicolumn{1}{r|}{\textit{Order of Trials}:  $F(1,107) = 39.02, p < 0.001$ \textbf{\textcolor{red}{*}}} \\\rule{0pt}{2.5ex} 
         & \multicolumn{1}{r|}{\textit{Interaction Effect}: $F(1,107) = 39.02, p < 0.001$ \textbf{\textcolor{red}{*}}} \\  \cline{2-2} \rule{0pt}{2.5ex} 
         & Post-Hoc Test \\ \cline{2-2} \rule{0pt}{2.5ex} 
         & \multicolumn{1}{r|}{\textbf{\EC{}} vs. \EW{} ($p<0.001$) \textbf{\textcolor{red}{*}}} \\ \rule{0pt}{2.5ex} 
         & \multicolumn{1}{r|}{\textbf{\EC{}} vs. \NC{} ($p<0.001$) \textbf{\textcolor{red}{*}}} \\ \rule{0pt}{2.5ex} 
         & \multicolumn{1}{r|}{\textbf{\EC{}} vs. \NW{} ($p<0.001$) \textbf{\textcolor{red}{*}}} \\ 
         
         \hline
         
    \end{tabular}}
    \captionsetup{belowskip=-8pt}
    \caption{Summary of results for average trust and change of trust. 
    We used a two-way factorial ANOVA test for the main effect and a Tukey HSD test for pairwise comparison.
    For the post-hoc results, bold texts represent the conditions with (a) higher trust and (b) more change.}
    \label{table:trst}
\end{table}

\begin{figure}[!t]
\captionsetup{belowskip=-5pt}
    \begin{subfigure}{0.49\columnwidth}
        \captionsetup{justification=centering}
        \caption{Average Trust Rating}
        \centering{\includegraphics[width=\columnwidth]{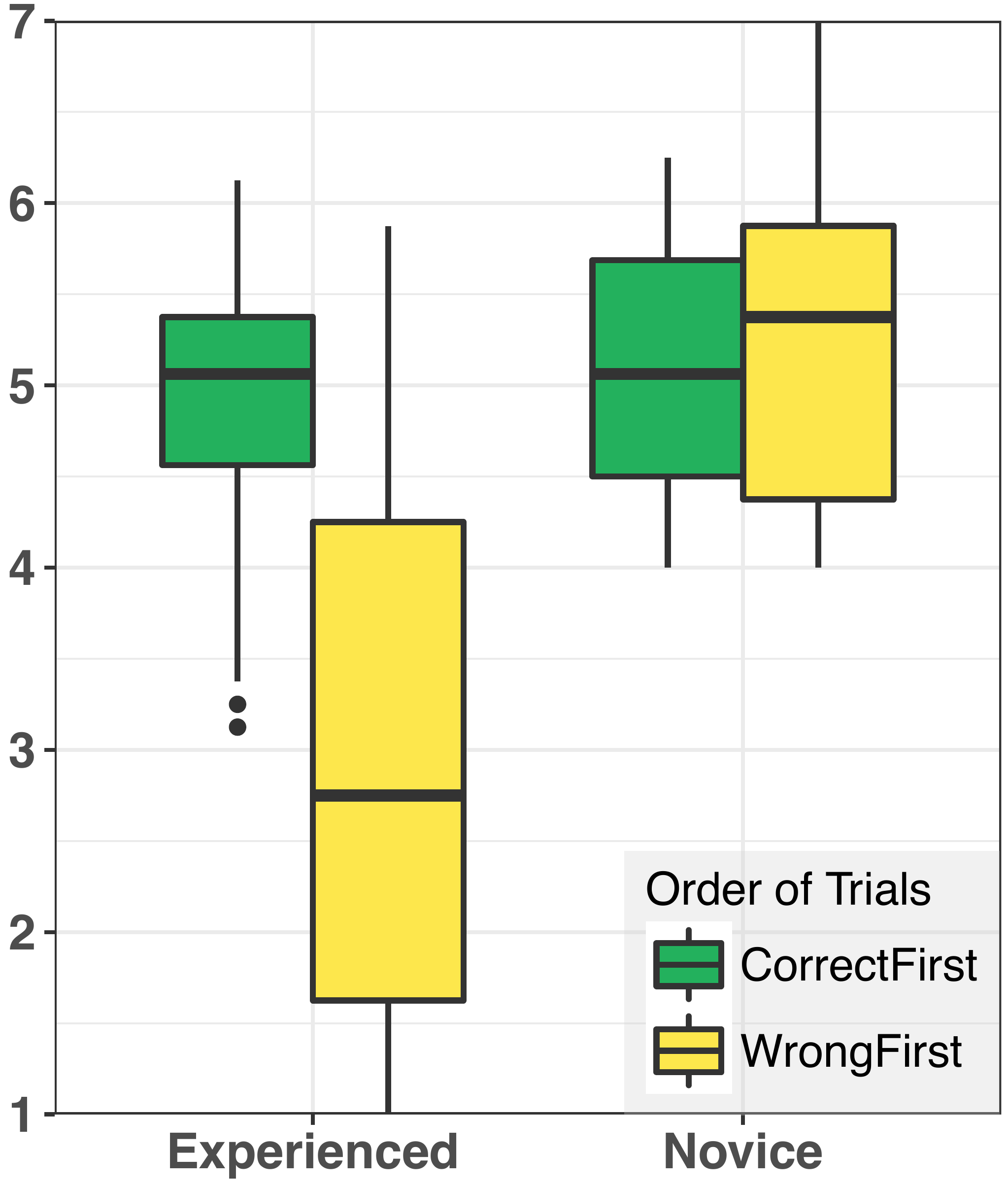}}
        \label{fig:trstAvg}
    \end{subfigure}
    \begin{subfigure}{0.49\columnwidth}
         \captionsetup{justification=centering}
        \caption{Change of Trust Rating}
        \centering{\includegraphics[width=\columnwidth]{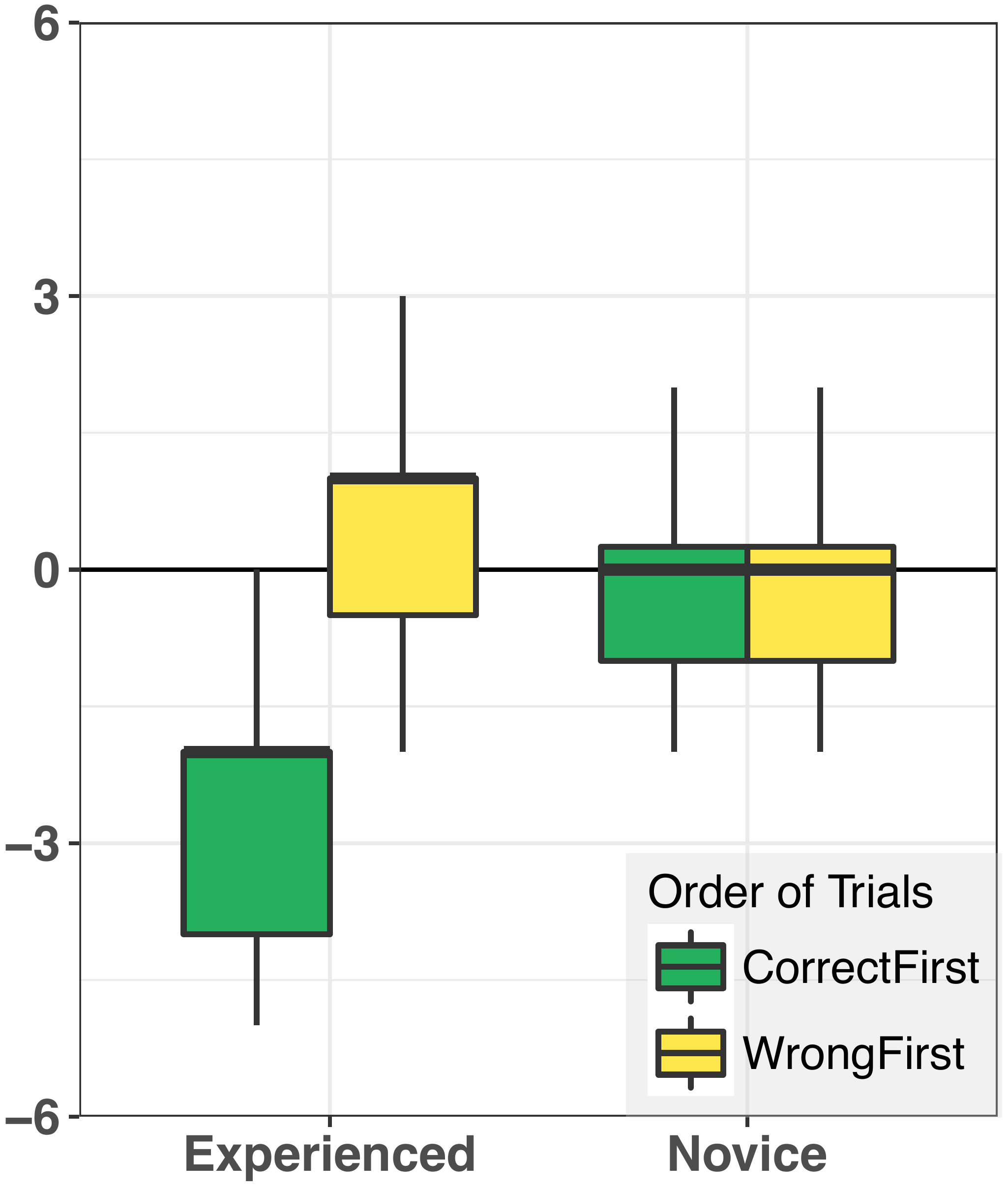}}
        \label{fig:trstChng}
    \end{subfigure}
    \caption{Results from self-reported trust in the \task{}. 
    (a) average of the 8 self-reported trusts, and (b) difference between the first and the last reported trust.
    Negative values indicate trust declining over time and positive points show increasing trust.}
    \label{fig:trst}
\end{figure}

We analyzed study results for the presented metrics based on the data collected from the \task{} and post-study questionnaire.
For simplicity, we use \NC{} and \NW{} condition names for \nov{} participants, as well as \EC{} and \EW{} condition names for \expert{} participants, with \corr{} and \wro{} order trials, accordingly.
For analysis, we used a two-way factorial ANOVA to test the main effects and Tukey HSD tests for posthoc pairwise comparisons. 
\begin{figure*}[t]
    \captionsetup{aboveskip=-5pt, belowskip=-5pt}
    \begin{subfigure}{1.05\columnwidth}
        \caption{Changes of trust over time for \nov{} participants.}
        \centering{\includegraphics[width=\columnwidth]{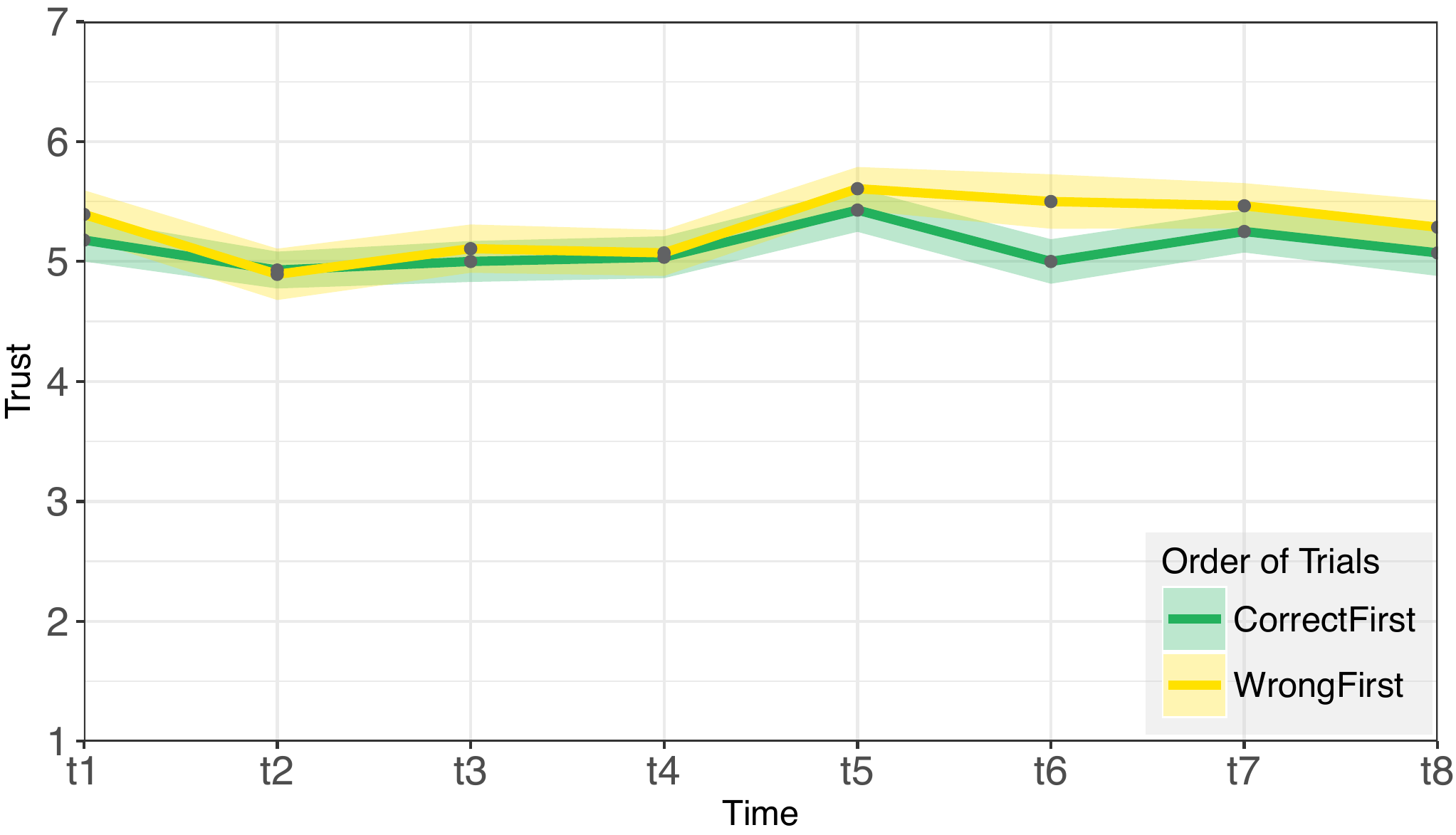}}
        \captionsetup{justification=centering}
        \label{fig:noviceChange}
    \end{subfigure}
    \begin{subfigure}{1.05\columnwidth}
        \caption{Changes of trust over time for \expert{} participants.}
        \centering{\includegraphics[width=\columnwidth]{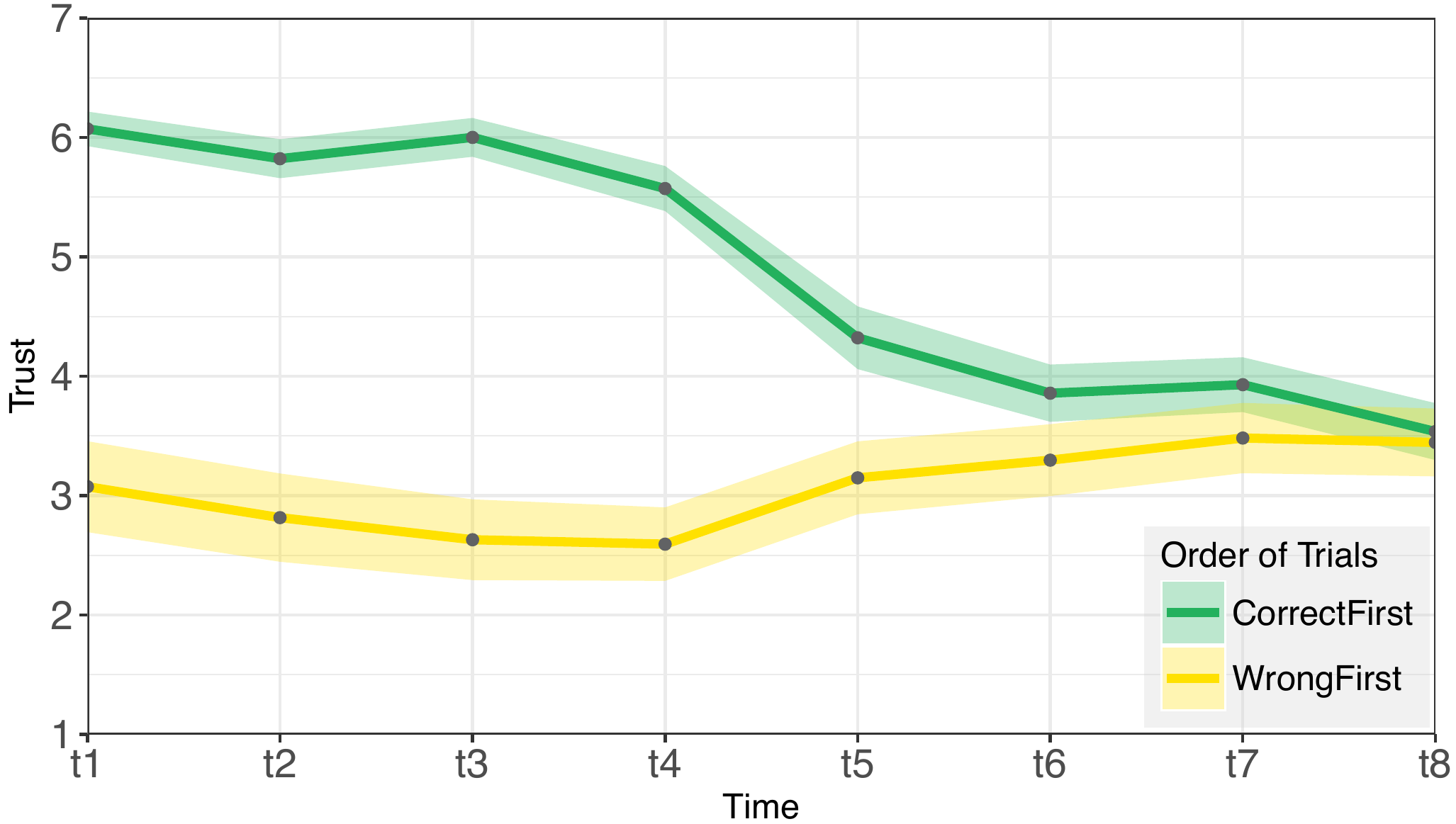}}
        \captionsetup{justification=centering}
        \label{fig:expertChange}
    \end{subfigure}
    \caption{Average of participants' self-reported trust after every 5 trials in the main task. The y-axis indicates level of trust, where 1 indicates distrust and 7 indicates total trust. The ribbon around each line shows the standard error of the mean. The x-axis shows the time-step when trust question is asked (every 5 trials).}
    \label{fig:trstOverTime}
\end{figure*}
\subsection{Data Pre-processing}

For quality verification, we removed the results from five participants due to data collection errors or evidence of lack of appropriate attention judged by their responses to the final open-ended questions.
This left us with data from 111 participants, with \NC{}, \NW{}, \EC{}, and \EW{} having 28, 28, 28, and 27 data points, accordingly.

\subsection{Average Self-reported Trust}

We tested the effects of background domain knowledge and order of observing system correctness on user self-reported trust.
Participants rated their trust in the system eight times during the \task{}.
To address our first two hypotheses (H1 and H2), we calculated the average trust for each participant and compared them to find differences across the conditions.
Table~\ref{table:trst}a and Figure~\ref{fig:trst}a show the summary of the results and distribution of this data.
The findings provide strong evidence that the average trust was significantly affected by domain knowledge and order of observing correct outputs.
However, a significant interaction effect indicates these factors are interdependent.
The pairwise comparison reveals that the order of observing correct system outputs is only significantly affecting trust for knowledgeable participants, which aligns with H1.
Moreover, users with domain expertise and positive first impressions report significantly higher trust compared to those with negative first impressions (H2).

\subsection{Changes in Trust over Time}

To test our third hypothesis (H3), participants rated their trust throughout the study so we could track how it evolves over time.
For each condition, we calculated the average trust of all participants per time-step, resulting in one trust value at each of the eight time-steps.
Figures~\ref{fig:noviceChange} and ~\ref{fig:expertChange} show line charts of changes over time for the \nov{}s and the \expert{} groups, respectively.

\begin{table}[t!]
    \resizebox{\columnwidth}{!}{
    \def\arraystretch{1.4}
    \begin{tabular}{|c|c|c|c|}
    \hline 
        \multirow{2}{4em}{Ordering Condition} & \multicolumn{3}{c|}{Common Themes}\\ \cline{2-4}
         & \textbf{Strong Distrust} & \textbf{Trust Decreased} & \textbf{Trust Increased} \\ \hline
         \textbf{CorrectFirst} & 0 & 21 & 0 \\ \hline
         \textbf{WrongFirst} & 11 & 5 & 8 \\ 
         \hline
    \end{tabular}}
    \captionsetup{belowskip=-10pt}
    \caption{Most common qualitative themes identified for how trust changed over time for participants from the experienced condition.
    The themes were retrieved from an open-ended question at the end of the study where participants were asked how their overall trust changed over time.
    }
    \label{tab:qual}
\end{table}

In order to statistically compare the magnitude and direction of change in trust over time across the conditions, we calculated the difference of initial trust and final trust.
With this measure, negative values indicate declining trust and positive values indicate increasing trust.
Table~\ref{table:trst}b and Figure~\ref{fig:trst}b show the results for this comparison.
Experienced participants in the \corr{} condition, had a significantly larger change-of-trust than those in the \wro{} condition.
The direction of this change was negative, indicating a loss of trust.
To understand these results further,  refer to Figure~\ref{fig:expertChange}.
Participants from the \corr{} condition start with higher trust, which decreases over time; this is expected as the system accuracy lowers with time.
In contrast, those from the \wro{} condition start off with lower trust due to their negative first impressions, and the magnitude of their change-in-trust is significantly less than their counterparts', slightly going up while remaining relatively low.
We did not observe any significant differences for novice participants.

We further reviewed the open-ended responses from the \expert{} participants to analyze how their trust in the system changed over time to understand the trends and themes based on first impressions.
A summary of the main observations is presented in Table~\ref{tab:qual}.
We expected \expert{} participants to have a proper understanding of how and when the system accuracy changed.
According to Yu et al.~\shortcite{yu2017trust}, this understanding should reflect in their change of trust.
Thus, we counted the number of participants whose comments indicated their assessments matched our expectation;
that is, a decrease in trust for \corr{} and an increase in trust for \wro{} participants.
In total, 21 out of 28 \expert{} participants in the \corr{} condition stated that their trust was high in the beginning but lowered over time.
From those in the \wro{} condition, only 8 out of 27 indicated a slight increase in their trust.
However, different themes were observed among these responses.
For example, one participant who detected a slight increase in accuracy noted:
\begin{quote}
    \textit{``Strangely, the system accuracy got much better towards the end, but by then I distrusted the system's [outputs, as] the misidentifications it made were too scandalous.''}
\end{quote}
Similarly, 11 out of 27 participants stated that regardless of the change in their trust, they did not trust the system at all.
For instance, one participant noted:
\begin{quote}
    \textit{``I didn't trust the system in the beginning and as the test continued, I only became surer that my distrust was the appropriate response.''}
\end{quote}
Moreover, 5 participants mentioned that their trust decreased over time, which is unexpected since the system grew more accurate towards the end.

These observations---backed by the statistical analysis for changes of trust---support our hypotheses (\textit{H1} and \textit{H3}) that first impressions of a system with local scope only matter when the user has background knowledge of the domain.
Positive first impressions provide a chance for users to build trust and not give up on the system when it makes mistakes, while negative first impressions can cause an overall distrust in the system.

\subsection{Post-Study Questionnaire}

\begin{figure}[!t]
\captionsetup{belowskip=-3pt}
    \begin{subfigure}{0.49\columnwidth}
        \captionsetup{justification=centering}
        \caption{Error of Perceived Accuracy (Percentage)}
        \centering{\includegraphics[width=\columnwidth]{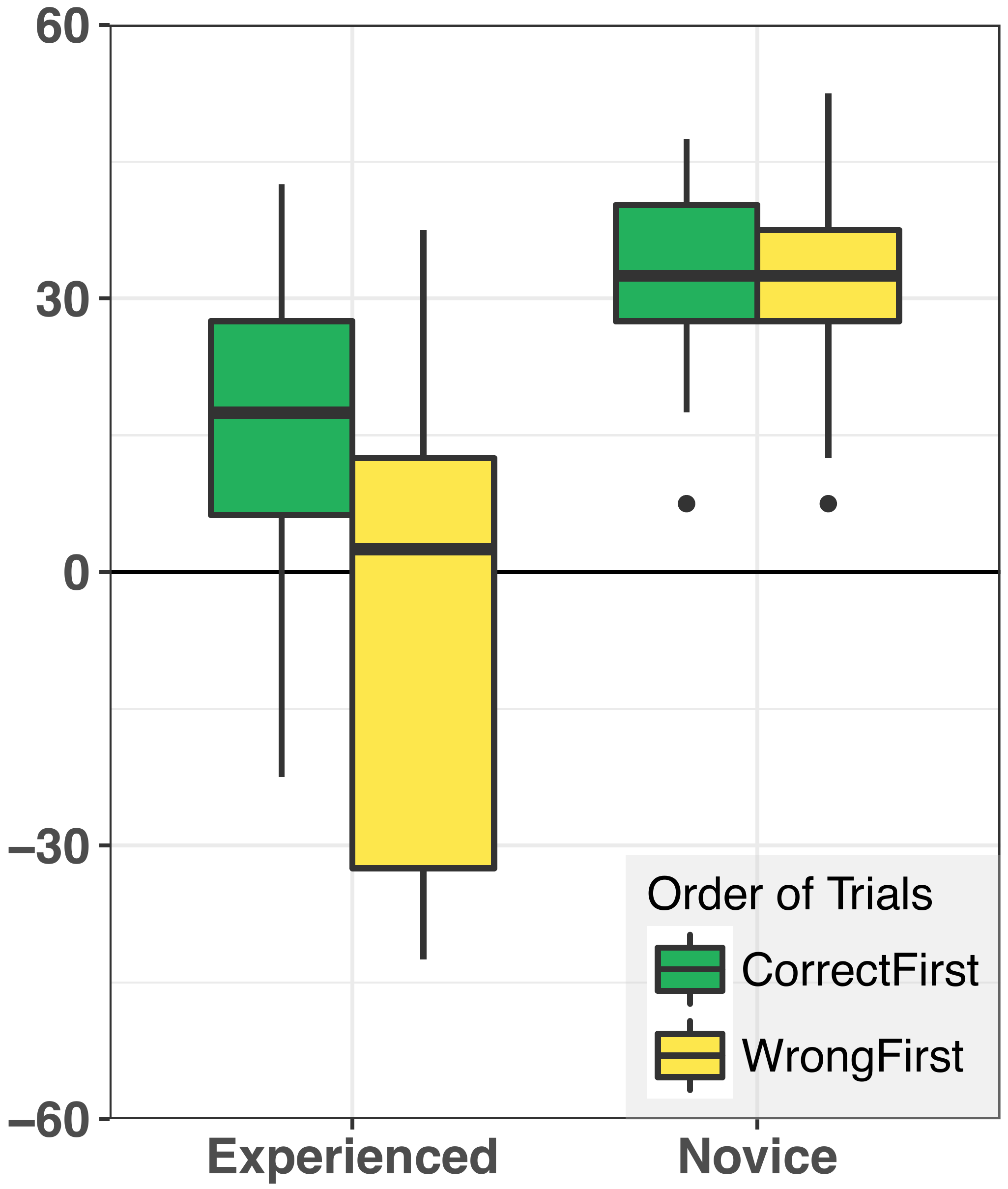}}
        \label{fig:accErr}
    \end{subfigure}
    \begin{subfigure}{0.49\columnwidth}
        \captionsetup{justification=centering}
        \caption{Average responses from Trust-in-XAI questionnaire}
        \centering{\includegraphics[width=\columnwidth]{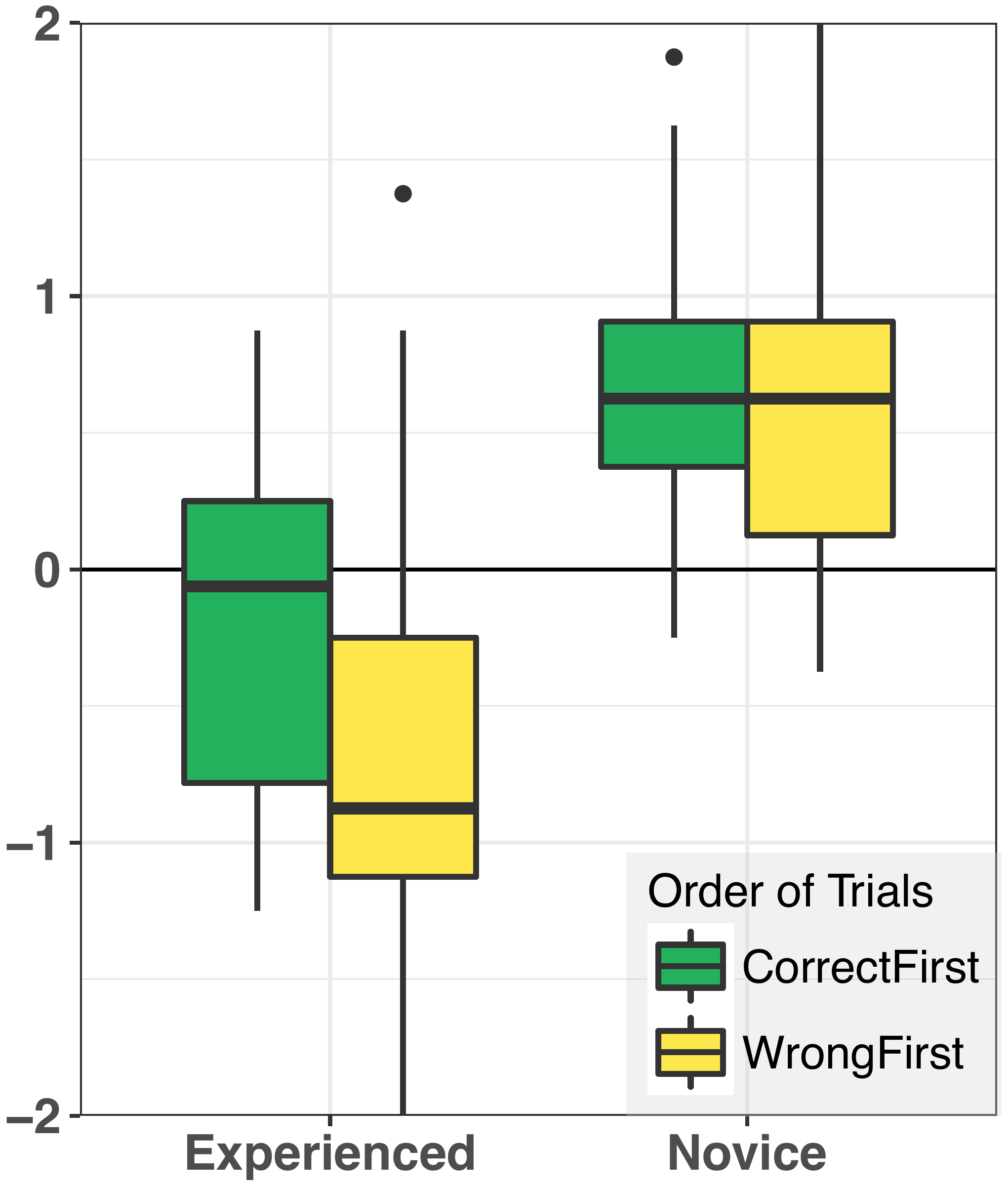}}
        \label{fig:trstQ}
    \end{subfigure}
    \caption{(a) Error of  user-estimation of system (difference between user estimations and the actual observed system accuracy). 
    Positive numbers represent overestimation and negative numbers represent underestimation of system accuracy.
    (b) Average of user responses to trust questionnaire results.
    Higher numbers indicate higher level of trust.}
    \label{fig:accAndTrst}
\end{figure}

After the \task{}, participants estimated the accuracy of the  classification system.
It is important to keep in mind that all participants (regardless of the condition) observed the same simulated classification results with the same controlled accuracy across observed instances.
The only difference was the order of observing the correct classifications.
We assessed the error of each participant's perceived system accuracy by calculating the difference between the estimated accuracy and actual observed system accuracy.
In addition, participants answered a set of questions about trust in automation and explainable AI systems~\cite{hoffman2018metrics} through a 5-point Likert scale, where higher values indicate higher trust in the intelligent system.
We calculated the average of all questionnaire responses into one single score per participant and analyzed them to test for differences among conditions.
Table~\ref{table:acc} and Figure~\ref{fig:accAndTrst} show a summary and distribution of the results.

The results show that novices significantly overestimated the system while experienced participants may underestimate or overestimate the accuracy based on the order of observing the correct trials.
The results from the trust questionnaire demonstrate that novice participants tended to trust the system significantly more than those with domain experience, which aligns with our first hypothesis (\textit{H1}).

\begin{table}[!t]
    \resizebox{\columnwidth}{!}{
    \begin{tabular}{|c|c|}
         \hline
         \rule{0pt}{2ex}
         \multirow{6}{*}{\textbf{(a) Error of Perceived Accuracy}} & Main Effect\\ \cline{2-2} \rule{0pt}{2.5ex} 
         & \multicolumn{1}{r|}{\textit{Domain Knowledge}: $F(1,107) = 76.88, p < 0.001$ \textbf{\textcolor{red}{*}}} \\ \rule{0pt}{2.5ex} 
         & \multicolumn{1}{r|}{\textit{Order of Trials}:  $F(1,107) = 14.48, p < 0.001$ \textbf{\textcolor{red}{*}}} \\\rule{0pt}{2.5ex} 
         & \multicolumn{1}{r|}{\textit{Interaction Effect}: $F(1,107) = 13.13, p < 0.001$ \textbf{\textcolor{red}{*}}} \\  \cline{2-2} \rule{0pt}{2.5ex} 
         & Post-Hoc Test \\ \cline{2-2} \rule{0pt}{2.5ex} 
         & \multicolumn{1}{r|}{\textbf{\EC{}} vs. \EW{} ($p<0.001$) \textbf{\textcolor{red}{*}}} \\
         \hline
         \hline
         \rule{0pt}{2ex}
         \multirow{4}{*}{\textbf{(b) Trust in XAI questionnaire}} & Main Effect\\ \cline{2-2} \rule{0pt}{2.5ex} 
         & \multicolumn{1}{r|}{\textit{Domain Knowledge}: $F(1,107) = 87.84, p < 0.001$ \textbf{\textcolor{red}{*}}} \\ \rule{0pt}{2.5ex} 
         & \multicolumn{1}{r|}{\textit{Order of Trials}:  $F(1,107) = 3.75, p = 0.055$ \small(NS)} \\\rule{0pt}{2.5ex} 
         & \multicolumn{1}{r|}{\textit{Interaction Effect}: $F(1,107) = 2.10, p = 0.149$ \small(NS)} \\ 
         
         \hline
         
    \end{tabular}}
    \caption{Summary of results for error of perceived accuracy and trust questionnaire. We used a two-way factorial ANOVA to test the main effect and a Tukey HSD test for pairwise comparison.
    (a) for the post-hoc results, the condition with bold text shows higher overestimation of accuracy.}
    \label{table:acc}
\end{table}

\section{Discussion}
Overall, our results demonstrate the importance of first impression formation with users with domain expertise and how it affects their trust.
In this section, we discuss the results more generally, their importance to system designers, and possible future directions.
\subsection{Interpreting the Results}

Our first goal in this study was to understand whether first impression formation is influenced by domain expertise.
Different implicit and explicit trust measures clearly indicate that novice users over-trusted the intelligent system although its overall accuracy was low.
However, since domain experienced users tend to be more skeptical due to their knowledge, their overall trust depended on their early observations of the system performance.
Domain experts' perception of system accuracy also varied by these first impressions, while novices always overestimated the accuracy.
Previous research by Papenmeier et al.~\shortcite{papenmeier2019model} demonstrated users cannot be tricked into trusting a low accuracy intelligent system with high fidelity explanations.
Our work builds on their findings and shows that given a domain-specific task where domain knowledge might be beneficial, novice users are prone to trusting such systems as they do not have enough knowledge to detect errors.

When comparing changes of user trust over time, domain experienced users showed different trends of trusting the system depending on the order of observed errors.
Experienced users with positive first impressions had a significantly higher magnitude of change in their trust compared to those with a negative first impression (Figure~\ref{fig:trst}b).
This observation indicates that with positive first impressions, expert users tend to adjust their trust, whereas those with negative first impressions start with lower trust which stays low throughout the usage.

As previous work in automation bias and psychology shows, it is easier to lose trust than to reestablish trust~\cite{hoffman2013trust}.
Starting with a system with good performance, experts are likely to form trust initially and to adjust their trust over time---also known as swift trust~\cite{hoffman2013trust}.
However, our results show that expert users with negative first impressions lose trust in the system and stated that they are not willing to use it in the future (see Table~\ref{tab:qual}).
This has implications on real-world systems, as users might not continue using a system they do not trust~\cite{dietvorst2015algorithm}.

\subsection{Implications for Intelligent System Designers}

This study presents important findings for intelligent system designers who are involved with designing domain-specific systems with various target users.
Designing one system for all users is indeed tricky and requires certain considerations with user domain knowledge.
Failing to account for such considerations can cause various problems such as under-reliance and over-reliance.
Our results indicate that while experienced domain users tend to be more skeptical of the system, novices might not be able to catch system problems and suffer from automation bias.

System designers can incorporate techniques to help fill the knowledge gap for novice users and utilize techniques to guide domain experts into observing system performance, e.g., by using a more detailed explanation interface to provide more information.
Rather than allowing users to use a system with zero understanding or a poor mental model of how the system works, designers can incorporate introductory sessions to alert users about system strengths and shortcomings so that users can decide whether and when they should trust the system's outputs.
Alternatively, designers can provide a high-level overview of the model with key information (e.g., system accuracy and known weaknesses) that can influence impression formation.
Additionally, one approach for consideration could be showcasing examples of both correct and incorrect predictions or outcomes in the beginning of usage to help experts develop first impressions, covering the variability of system capabilities to reduce the risk of encountering an unrepresentative sample by chance.
Further research would need to explore the implications of such an approach.
The showcasing method could also consider attempts to more strongly encourage users to review explanations for both correct and incorrect examples.
Our results indicate that users tend to check the explanation for further information when they encounter errors, but not necessarily when they perceive the system to be correct.
For novice users, however, errors need to be shown and explained to circumvent automation bias.

\subsection{Limitations and Future Work Opportunities}

This study contributed novel findings regarding how domain expertise can affect  first impression formations and user trust.
Our results show that novice users trusted the system regardless of the order of observing system errors and the overall low system accuracy.
One possible explanation for this observation is novice users' inability to identify errors.
A user's assigned level of expertise might vary with the domain and task at hand.
While some systems consider novices as students or inexperienced domain knowledgeable users, we expected novices to have little or no domain knowledge at all.
Thus, our study results from domain experts might also be observed for novices if they have the ability to judge system errors correctly.

Measuring trust is tricky, and any chosen evaluation methodology will have limitations.
Directly asking subjects to rate their trust might bias them about the purposes of the study and hence, affect their response.
Another issue relates to the use of Likert-scales for self-reported trust estimations and whether participants are able to differentiate the values in their response.
Although 7-point Likert-scales are generally reliable for ordinal self-reported measures~\cite{oaster1989number}, we cannot control nor can we precisely know how each participant differentiates each point (e.g., value 5 from 6).
In our study, we looked to qualitative data and free-response questions to help address this limitation and provide a better understanding for the collected explicit quantitative trust metrics.
While the qualitative and quantitative results align, our study is limited in its ability to dissect specific characteristics of the observed trust and mistrust due to the open-ended nature of the free response data collection.

To maintain experimental control for our study, we selected a task where there is a definition for correct predictions.
In other words, for these tasks, there is a concrete distinction between when the system makes a mistake and when not.
To achieve this, the study is based on a multi-class image classifier with visual explanations.
While such tasks are quite common in different domains and our results can be generalized for such intelligent systems, future work is required to verify if these findings hold for more exploratory and complex tasks.
Specifically, for these tasks, errors might be challenging for users to detect, while they can also be difficult or impossible to define.
For example, missing information or missing values can cause a model to predict different outcomes, all of which may be correct based on different hypothetical values for the missing information.
As another example, recommendation systems strive to help users by making suggestions, but how these suggestions fit a user's needs is not easy to assess, and ``false suggestions'' are often impossible or difficult to define.
Future research can extend the study of our research questions to such alternative analytical systems and tasks.

Finally, our findings show strong impression-formations for expert users based on instance-level observations of system performance.
The presented study used a system with simple representations of model outputs.
Future research of higher-level representations or visualizations can investigate how our findings generalize to contexts that allow deeper expert analysis of the model as-a-whole.

\section{Conclusion}
In this paper, we present a controlled human experiment to understand how user domain knowledge can affect first impression formation and trust calibration over time.
Choosing entomology as our domain, we recruited domain-knowledgeable and novice participants to review outputs of a simulated arthropod-classification system with a low accuracy.
Through the course of the study, we asked the participants to rate their trust in the system, and in the end, we measured overall trust implicitly and explicitly.
Our significant results show that only those with domain knowledge form first impressions of the system.
With a low accuracy system, we expected low overall trust from the subjects.
However, Encountering errors early-on resulted in a lower trust over time and a reluctance to use and rely on the system in the future.
Though with positive first impressions, subjects calibrated their trust as they observed the system performance and were more likely to use the system in the future.

\section{Acknowledgments}
This work was supported by the DARPA Explainable Artificial Intelligence (XAI) Program under award number N66001-17-2-4032 and by NSF award 1900767.
We would like to thank Dr. Vincent Bindschaedler for providing constructive feedback and suggestions on this study, as well as Emma Drobina, Brianna Richardson, and Prashant Singh for their initial efforts on this project.

\bibliographystyle{aaai}
\bibliography{references}
\end{document}